\documentclass[final,3p,11pt,twocolumn,times]{elsarticle}

\usepackage{amsmath,amssymb}
\usepackage[table,xcdraw]{xcolor}
\usepackage{booktabs,multirow}
\usepackage{subcaption}
\usepackage{lineno}
\usepackage{comment}
\usepackage{bm}
\usepackage{upgreek}
\usepackage[section]{placeins}
\usepackage{algorithm2e}
\usepackage[colorlinks=true, pdfstartview=FitV, linkcolor=black, citecolor=black, urlcolor=black]{hyperref}

\usepackage{enumitem}

\setlist[enumerate,1]{label=\arabic*}
\setlist[enumerate,2]{label=\theenumi.\arabic*}
\setlist[enumerate,3]{label=\theenumii.\arabic*}
\makeatletter
\def\ps@pprintTitle{%
   \def\@oddhead{}%
   \let\@evenhead\@oddhead
   \def\@oddfoot{}%
   \let\@evenfoot\@oddfoot
}
\makeatother

\begin{document}

\begin{frontmatter}



\title{LGR-MPC: A user-friendly software based on Legendre-Gauss-Radau pseudo spectral method for solving Model Predictive Control problems}


\author[author-ise]{Saeid Bayat\corref{author-corr}}
\author[author-ise]{James T.Allison}

\affiliation[author-ise]{
    department = {Department of Industrial and Enterprise Systems Engineering,},
    organization = {University of Illinois at Urbana-Champaign,},
    city={Urbana},
    state={IL},
    country={USA}
}

\cortext[author-corr]{Corresponding author. Email: bayat2@illinois.edu, Address: 509 E. White Ave., \#1, Champaign, IL 61820.}

\begin{abstract}

Active components, such as actuators, constitute a fundamental aspect of engineering systems, affording the freedom to shape system behavior as desired. However, this capability necessitates energy consumption, primarily in the form of electricity. Thus, a trade-off emerges between energy usage and desired outcomes. While open-loop optimal control methods strive for efficiency, practical implementation is hampered by disturbances and model discrepancies, underscoring the need for closed-loop controllers. The Proportional-Integral-Derivative (PID) controller is widely favored in industry due to its simplicity, despite sub-optimal responses in many cases. To bridge this gap, Model Predictive Control (MPC) offers a solution, yet its complexity limits its broad applicability. This paper introduces user-friendly Python-based MPC software, enabling easy access to MPC. The effectiveness of this software is demonstrated through multiple examples, including those with a known analytical solution

\end{abstract}





\end{frontmatter}


\section{Introduction}
\label{sec:intro}
Actuators, known as active components, are commonly employed in engineering systems to fulfill the objectives set by the designer. By manipulating the actuators, the system's response can be fine-tuned to achieve optimal outcomes and stabilize unstable systems. However, the operation of these components necessitates energy consumption, resulting in a trade-off between the actions performed and energy usage. Designers strive to attain the best possible results while minimizing energy consumption. To accomplish this objective, Open Loop Control (OLC) problems are formulated, where no structural assumptions are made, enabling the optimization algorithm to generate the optimal control signal freely. While OLC offers advantages, it faces a significant limitation: it cannot be practically implemented due to the disparities between the real system and the model used for OLC design \citep{chen2020robust}. Thus, the adoption of Closed Loop Control (CLC) becomes necessary, utilizing system state feedback. CLC controllers excel at handling disturbances, model mismatches, measurement noise, and other factors, making them widely applied across various applications. 

The industry widely recognizes Proportional-Integral-Derivative (PID) control as the most renowned CLC method, thanks to its favorable response and straightforward design \citep{foley2005comparison}. However, the PID response may significantly differ from that of OLC because PID has a specific structure that prevents it from replicating an OLC-like response, even with the assistance of optimization algorithms. This disparity between OLC and CLC can be bridged by Model Predictive Control (MPC). MPC offers varying levels of information, allowing for the inclusion of controllers ranging from those with no knowledge of future events to those equipped with complete information. Despite its advantages, MPC faces a significant drawback in terms of computational cost \citep{leuer2014real,bayat2023ss}. However, recent advancements in transistor and processor design have expanded the applicability of MPC in engineering systems \citep{frison2015mpc,bayat2023ss,SaeidBayat-Vehicle}. Another challenge lies in the design of MPC itself, particularly for industry professionals whose expertise may not primarily lie in control theory. To address this, user-friendly software becomes essential in making MPC a more versatile controller. This paper introduces a solution where users are not required to possess in-depth knowledge of the MPC design process. By leveraging MPC hyper parameters, users can effectively design an MPC for their specific study system without delving into intricate details.

The concept of developing closed-loop control based on an open-loop controller was first introduced in literature more than 50 years ago. Notably, \citet{lee1967foundations} presented the essence of Model Predictive Control (MPC). In the present day, with advancements in transistor technology and the development of high-frequency processors, MPC finds applications in a wide range of areas. These include suspension systems \citep{giorgetti2006hybrid}, automotive powertrains \citep{saerens2008model}, power converters \citep{richter2010high}, vehicle traction control \citep{borrelli2006mpc}, and various other domains.

Several frameworks have been introduced in the literature to address Model Predictive Control (MPC) challenges. Notable examples include DO-MPC \citep{lucia2017rapid}, ACADO \citep{houska2011acado}, and MUSCOD II \citep{kuhl2007muscod}. While these tools offer solutions, many of them demand specific software dependencies or necessitate defining problems using particular syntax, which might not be user-friendly. For instance, in the case of MUSCOD II \citep{kuhl2007muscod}, users are required to submit files without access to the underlying code, which is also not publicly available. On the other hand, DO-MPC \citep{lucia2017rapid} provides an open-access modular codebase, allowing users to define and modify different modules as needed. However, the usage of CASADI \citep{andersson2019casadi} for automatic differentiation poses a challenge. Users must formulate problems using CASADI's syntax, which has a steep learning curve initially. Furthermore, DO-MPC \citep{lucia2017rapid} assumes the control horizon to be the same as the prediction horizon, limiting flexibility. Additionally, due to the CASADI dependency, working with black-box functions becomes difficult or even impractical. This is problematic because many engineering systems involve black-box functions in dynamics, constraints, or objective functions.

Several tools, including DO-MPC \citep{lucia2017rapid}, have publicly available codebases. However, the complexity of these codes hinders their educational utility, making it challenging to comprehend the functioning of each component for teaching purposes. Moreover, certain techniques like single shooting employed in tools like SS-MPC \citep{bayat2023ss} suffer from issues such as ill-conditioning. Furthermore, specific MPC toolsets are tailored for particular tasks like building operations. Examples of such task-specific software include MShoot \citep{arendt2019mshoot}, MPCPY \citep{blum2019mpcpy}, and TACO \citep{jorissen2019taco}, all designed exclusively for this purpose.

To address the aforementioned challenges, a novel MPC toolset is introduced in this paper, characterized by the following key attributes:
\begin{itemize}
    \item Open Source: The toolset is made publicly available, with access to the underlying source code. This encourages broader utilization, including educational purposes, and provides transparency.
    \item Independence from Third-Party Software: This toolset operates autonomously, without necessitating the installation of additional third-party software. This alleviates concerns about compatibility across different machines.
    \item Simplified Syntax: Unlike tools relying on specific syntax like CASADI \citep{andersson2019casadi}, this toolset is developed in Python, requiring only a basic understanding of Python for effective usage.
    \item Black Box Function Compatibility: Given that this paper does not rely on automatic differentiation, it becomes possible to employ any black box function seamlessly. Therefore, users are not compelled to either rephrase their black box functions using a predefined syntax or modify their problem to convert the black box function into a symbolic form.
    \item Stability through Pseudospectral Methods: Employing pseudospectral methods, the toolset transforms the problem into a nonlinear program, mitigating the ill-conditioning problems associated with certain methods like Shooting \citep{bayat2023ss}.
    \item Comprehensibility: The paper elaborates on the essential steps involved in formulating MPC as a Nonlinear Programming (NLP) problem. This offers a detailed understanding of the toolset's inner workings.
    \item Flexibility: Each aspect, including dynamics, objectives, constraints, MPC parameters, and optimization settings, can be modified easily without necessitating the complete redefinition of the problem. This flexibility empowers users to assess the impact of each parameter on the solution. Additionally, control horizon and prediction horizon can be defined independently, offering greater control in MPC design compared to tools like DO-MPC \citep{lucia2017rapid}.
    \item Simplified Design: The toolset eliminates the need for users to manually implement the MPC loop and individual function calls for tasks such as simulation and result retrieval. The code automates these steps, providing users with a vectorized result across the entire time horizon.
    \item Active GitHub Repository: The toolset is backed by an active GitHub repository, ensuring ongoing updates and support for addressing reported issues.
    \item Accessible Complexity: Although toolsets like DO-MPC \citep{lucia2017rapid} offer modular frameworks applicable in various scenarios, such as robust MPC or state estimation, their intricacy often obscures the understanding of how distinct code components operate. In contrast, this toolset aims to strike a balance. It endeavors to provide software that remains relatively uncomplicated while establishing a robust foundation built on advanced techniques, such as pseudospectral methods, for designing MPC problems.
\end{itemize}

It's essential to note that this toolset is not intended to replace existing powerful MPC toolsets. Instead, its purpose is to specifically address the outlined goals, offering a solution that aligns with the aforementioned attributes. The remainder of this article is as follows: In Sec.~\ref{sec:MPC} through \ref{sec:Pseudo-spectral}, you will find a comprehensive overview of MPC parameters and Pseudo-Spectral methods. In Sec.~\ref{sec:Imp} the MPC implementation is discussed. Software structure is discussed in Sec.~\ref{sec:Sof-Struct} and different examples are solved in Sec.~\ref{sec:Examples}. Finally, Sec.~\ref{sec:Concl} provides a concluding summary. In \ref{sec:Methods}, a variety of methods for resolving MPC problems are discussed, and \ref{sec:Examples-appendix} contains three distinct examples that have been solved using this software.
\subsection{MPC}
\label{sec:MPC}

To design an MPC controller, three hyper parameters are needed:
\begin{itemize}
    \item Control Sampling Time ($T_s$): is the rate at which the controller executes the control algorithm. At the time interval between each sampling point, the control is constant and is equal to the previous time step. It also determines how fast the controller can react to disturbances and setpoint changes. In addition, it determines how advanced processors are needed because as $T_s$ decreases, the computation time increase, but with the benefit of responding faster to disturbances and setpoint changes. The rule of thumb to determine $T_s$ is to put 10 to 20 samples in system rise time, which is defined as the time that takes for the system step response to rise from $10 \%$ to $90 \%$ of the steady-state response \citep{bordons2020model}.
    
    \item Prediction Horizon ($p$): indicates the number of time steps utilized for making predictions using the system model. The product $pT_s$ represents the duration for which the prediction is made. Increasing the value of $p$ leads to more accurate MPC results as it provides more information about the system. However, this comes at the expense of higher computation costs. To strike the right balance, it is advisable to include sufficient samples to cover the system's settling time. The settling time refers to the duration when the error between the system's step response and its steady-state becomes less than or equal to $2\%$ \citep{wang2009model}.
    
    \item Control Horizon ($m$): $m \le p$ shows the number of time steps for which MPC calculates optimal actions. The maximum value for $m$ is $p$, and the length of the time window in which control is not constant is $mT_s$. In the time interval $[mT_s\,\,pT_s]$ control is constant and is equal to control action at time $mT_s$. A general rule to set $m$ is to use a value between $0.1 p$ and $0.2 p$ with a minimum of 3 \citep{allgower2012nonlinear}. 
\end{itemize}

MPC addresses a problem that is similar to optimal control in an open-loop setting but with an iterative approach. In each iteration of MPC, the controller solves an optimization problem to determine the optimal control sequence for a finite time horizon. This includes predicting the system's future behavior based on a model and optimizing the control inputs to minimize a specified cost function. The iterative nature of MPC allows for the incorporation of real-time measurements and the ability to adapt the control actions based on the evolving system dynamics, disturbances, and changing setpoints. By continuously updating the control inputs, MPC effectively addresses the dynamic nature of the system, leading to improved performance and robustness compared to open-loop optimal control. The general optimal control problem is shown in Eq.~\ref{eq: Ol_formula} \citep{herber2017advances,herber2014dynamic}. Here $t$ is time, $t_0$ is initial time, $t_f$ is final time, $u$ is action, $\bm \xi$ shows state, $\Phi$ is Mayer cost, $\mathcal{L}$ is Lagrange cost, $\mathcal{\bm C}$ shows path constraint and $\mathcal{\bm \phi}$ shows boundary constraint. Dynamis is also shown by $\bm f$. It should be noted that depending on the method used to solve the optimal control problem, we have a different notation for dynamic in the OLC formula. This will be discussed in the next section.

\begin{align}
    \label{eq: Ol_formula}
    \min_{\bm u(t)}\,\,\, &\Phi\left( \bm \xi(t_0), t_0, \bm \xi(t_f), t_f \right)+ \int^{t_f}_{t_0} \mathcal{L}\left(t,\bm\xi(t),\bm u(t) \right)dt \\
    &\mathrm{Subject\,\, to:}\nonumber \\
    &\dot{\bm \xi}-\bm f\left(t,\bm \xi(t),\bm u(t)\right)=0\nonumber \\
    &\mathcal{\bm C}\left( t,\bm\xi(t),\bm u(t) \right)\le0\nonumber \\
    &\mathcal{\bm \phi} \left(t_0,\bm \xi(t_0),t_f,\bm \xi(t_f) \right)= 0\nonumber 
\end{align}

\begin{figure*}[ht!]
    \centering
    \includegraphics[scale=0.6]{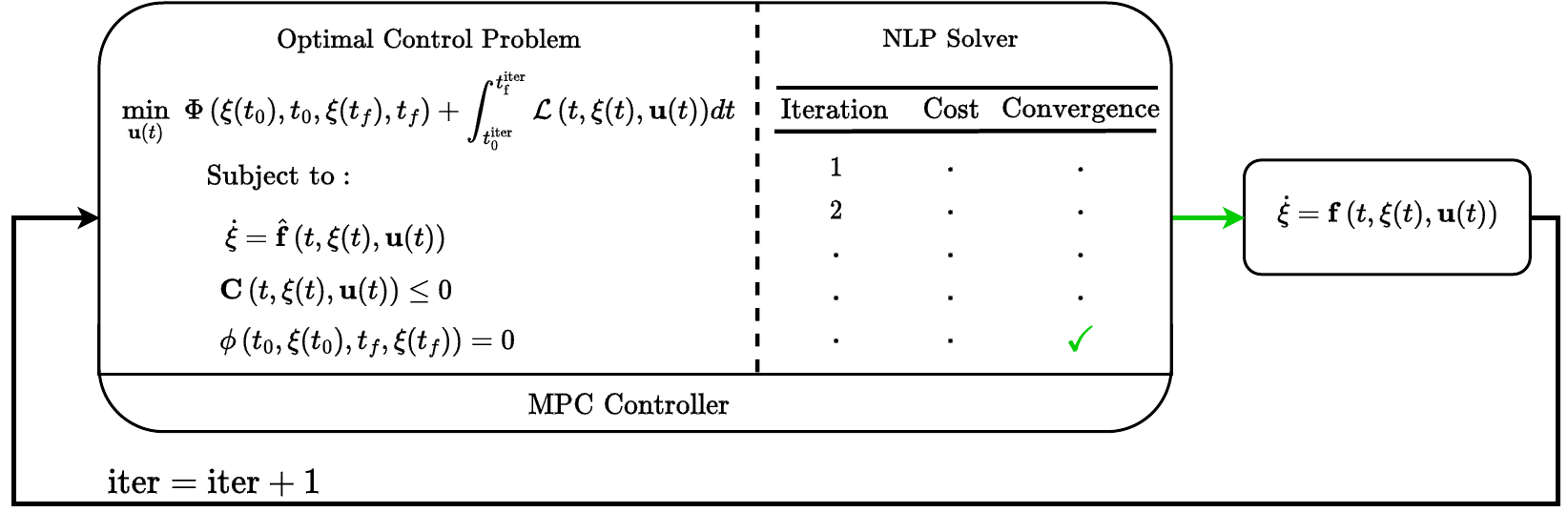}
    \caption{Schematic illustrating MPC loop. At each iteration, the optimization problem is solved using a nonlinear programming software while satisfying constraints and system model dynamics ($\hat {\bm{f}}$). When it converges, the first time step of the controller is applied to the real plant ($\bm f$).}
    \label{fig:MPC_loop}
\end{figure*}

The distinction between MPC and open-loop optimal control problems lies in the iterative nature of MPC. In each iteration, MPC solves a problem defined by Eq.~\ref{eq: Ol_formula}, where $t_0$ and $t_f$ represent the starting and final times for that particular iteration. After each iteration, the starting time moves forward by one sampling time $T_s$, while the final time is adjusted to ensure that the time interval between the initial and final times covers the entire prediction horizon. The MPC loop, illustrated in Fig.~\ref{fig:MPC_loop}, demonstrates how each iteration minimizes the objective over the corresponding time window while satisfying dynamic equations, path constraints, and boundary conditions. Once convergence is achieved, the resulting control actions are applied to the real system, denoted as $\bm f$, and the iteration count is increased by 1. It is important to note that the Mayer cost and boundary conditions are only considered when the initial or final time of the system falls within the time window being solved by MPC. During each iteration, an optimization algorithm is employed to solve the MPC problem. The model used in the optimization process, denoted as $\hat{\bm f}$, can differ from the actual system model. For instance, a simplified model might be used for faster computation, or disturbances may be unknown, resulting in the optimization model not accounting for them.

MPC problem shares similarities with the open-loop optimal control problem. \ref{sec:Methods} delves into various methods employed to address open-loop optimal control problems. This paper uses the pseudo-spectral method to solve the MPC problems. Pseudo-spectral methods can be divided into different categories depending on the nodes used for collocation and regression. This method is discussed in detail in Sec.~\ref{sec:Pseudo-spectral}.

\subsection{Pseudo spectral Methods}
\label{sec:Pseudo-spectral}
In pseudospectral methods, as discussed in \ref{sec:Methods}, the states and controls are discretized. High-order polynomials are employed to approximate the continuous function of states, while no assumptions are made for control signals at points other than the discretized ones. Collocation is used in the pseudospectral method to satisfy the dynamic equation. Specifically, a set of points, denoted as $\tau={\tau_1,\tau_2,\cdots,\tau_N}$, is employed to equate the left-hand side of the dynamic equation to the right-hand side. This results in the generation of defect constraints, represented by $\zeta$:
\begin{align}
    \dot{\bm \xi}(\tau_j)-\bm f(\tau_j,\bm \xi(\tau_j), \bm u(\tau_j))=0\,\,, (j=1,\cdots,N)
\end{align}

In pseudo-spectral methods, the collocation points ($\tau$) includes 3 main categories: Gauss methods, Radau method and Lobatto methods. In a gauss method, neither of the end points ($\tau_1,\,\tau_N$) are collocation points. In Radao method at least on of the end points ($\tau_1,\,\mathrm{or},\,\tau_N$) is a collocation point. In a Lobatto method, both end points ($\tau_1,\,\mathrm{and},\,\tau_N$) are collocation points.

In the pseudo-spectral method, the set of points used for collocation is obtained from the root of the Chebyshev or Legendre polynomial \citep{rao2009survey}. These points provide orthogonal collocation property. The benefit of this is that the quadrature approximation to a definite integral is pretty accurate \citep{rao2009survey}. When roots of Legendre polynomial are used, these methods are called Legendre-Gauss(LG), Legendre-Gauss-Radau(LGR), and Legendre-Gauss-Lobatto (LGL). All these points are defined on the interval $[-1,+1]$, So the time domain $[t_0,t_f]$ is mapped to $[-1,1]$ by using the following equation:
\begin{align}
    t=\frac{t_f-f_0}{2}\tau+\frac{t_f+t_0}{2}
\end{align}

\begin{figure}
\hspace*{-1.5cm}
    \centering
    \includegraphics[scale=0.5]{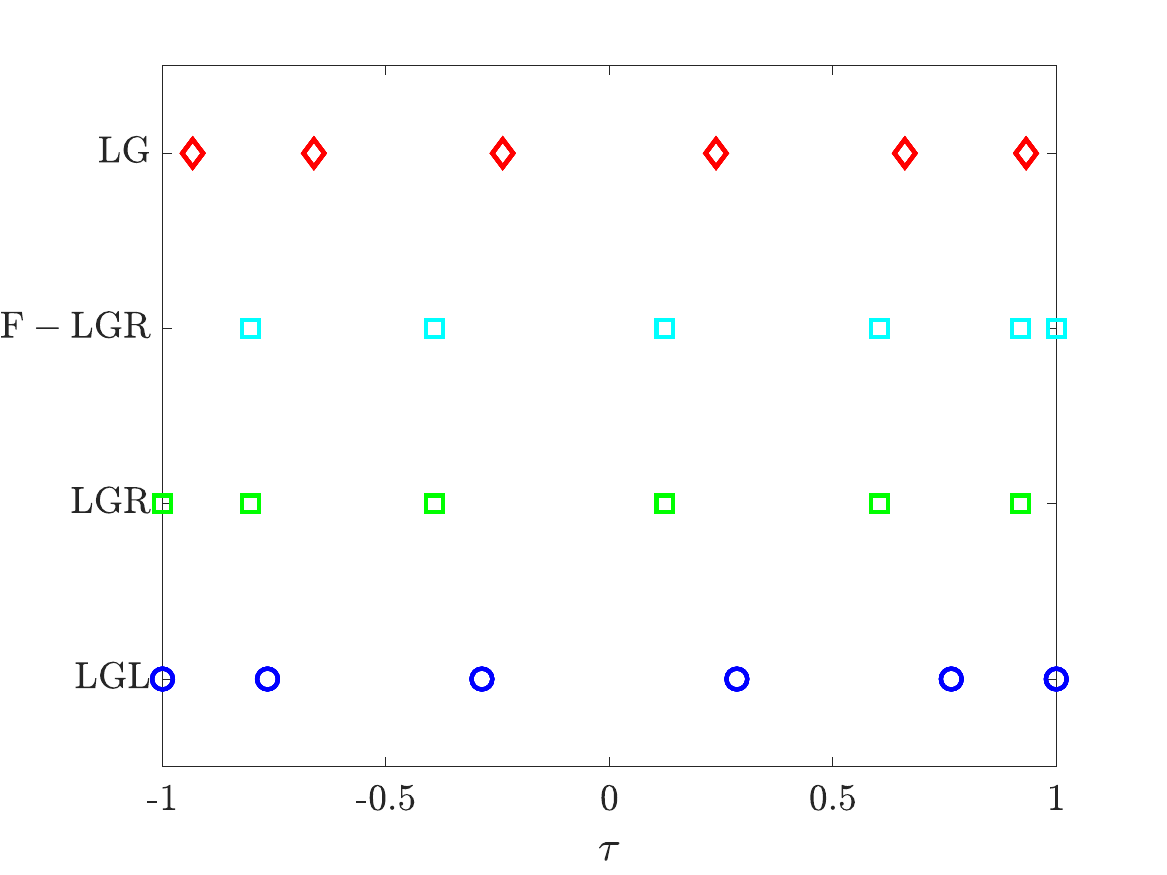}
    \caption{Collocation nodes of different pseudo-spectral methods}
    \label{fig:Pseudo_Nodes}
\end{figure}

In these method, the $N$ collocation points are obtained from Eq.~\ref{eq: nodes}. Figure~\ref{fig:Pseudo_Nodes} also shown the location of these points. As we see, LGL contains both endpoints, LGR contains just starting point, Flipped LGR includes end point, and LG contains no end point.
\begin{align}
\label{eq: nodes}
    &\mathrm{LG}: \,\, \mathrm{Roots\,\,of\,\,}P_N(\tau)\\
    &\mathrm{LGR}: \,\, \mathrm{Roots\,\,of\,\,}P_{N-1}(\tau)+P_{N}(\tau)\nonumber\\
    &\mathrm{LGL}: \,\, \mathrm{Roots\,\,of\,\,}\dot{P}_{N-1}(\tau) + \{-1\}\nonumber\\
    &\mathrm{Flipped\,\,LGL}: \,\, \mathrm{Roots\,\,of}\dot{P}_{N-1}(\tau) + \{+1\}\nonumber
\end{align}

As mentioned earlier, the open-loop optimal control problem will be transcribed to an NLP using direct methods. As a result, The equivalence of the NLP and the OLC is essential. In \citet{rao2009survey}, it is shown that the LGL methods do not yield a complete equivalence, but LGR and LG methods do. In addition, because the LGR method has a collocation point at the start, it is better than the LG method because after solving the NLP, the optimal control value can be obtained at time 0. In contrast, if the LG method is used, there is no value at time 0 for state or control. Therefore, the LGR method is used in this paper.

The procedure of LGR method is as follows:
Consider N LGR points, $\tau=\{\tau_1,\cdots,\tau_N\}$, where $\tau_1=-1$, and $\tau_N<+1$. We define a new point at $\tau_{N+1}=+1$. Then, Lagrange polynomials with degree N are used to interpolate states at LGR points plus $\tau_{N+1}=+1$, so we have:
\begin{align}
    &L_i(t)=\prod_{k=1,\, k\neq i}^{N} \frac{t-t_k}{t_i-t_k}\\
    &\bm \xi (\tau) \approx \bm \Xi(\tau)=\sum_{i=1}^{N+1}\bm \Xi_i L_i(\tau)
\end{align}
where $\bm \Xi_i=\bm \Xi(\tau_i)$. As a result, each state is approximated with a Lagrange polynomial with degree N. The collocation points are the N LGR points and are used to provide defect constraints that make the left-hand side of the dynamic equal to the right-hand side. By taking the derivative of the above equation, we have \citep{fahroo2008advances}:
\begin{align}
    \bm \dot{\xi} (\tau) \approx \bm \dot{\Xi}(\tau)=\sum_{i=1}^{N+1}\bm \Xi_i \dot{L}_i(\tau)
\end{align}

By equating the derivative of state to the dynamic for the LGR points, we have \citep{fahroo2008advances}:

\begin{align}
    \sum_{i=1}^{N+1}\bm \Xi_i \dot{L}_i(\tau_k)&=\frac{t_f-t_0}{2}\bm f\left( \bm \Xi_k, \bm U_k, \tau, t_0,t_f \right) , \,\,(k=1,\cdots,N)\\
    \sum_{i=1}^{N+1} D_{ki}\bm \Xi_i&=\frac{t_f-t_0}{2}\bm f\left( \bm \Xi_k, \bm U_k, \tau, t_0,t_f \right), \,\,(D_{ki}=\dot{L}_i(\tau_k))
\end{align}

Where $\bm U_k=\bm U(\tau_k)$, and the $D_{ki}$ ,$(1 \le k\le N)\,,\, (1\le i \le N+1)$ is a $N \times (N+1)$ non-square matrix and is called differentiation matrix. The D matrix is non-square because approximation of the states is used at $N+1$ points, but only the LGR points ($N$ points) are used for collocation. Let $\bm \Xi ^{\mathrm{LGR}}$ and $\bm U^{\mathrm{LGR}}$ be defined as:
\begin{align}
    \bm \Xi^{\mathrm{LGR}}&=\begin{bmatrix}
    \bm \Xi_1\\
    \vdots\\
    \bm \Xi_{N+1}
    \end{bmatrix}_{(N+1)\times n_{\xi}}\\
    \bm U^{\mathrm{LGR}}&=\begin{bmatrix}
    \bm U_1\\
    \vdots\\
    \bm U_{N}
    \end{bmatrix}_{(N)\times n_{u}}
\end{align}

\begin{figure*}
    \centering
    \includegraphics[scale=0.8]{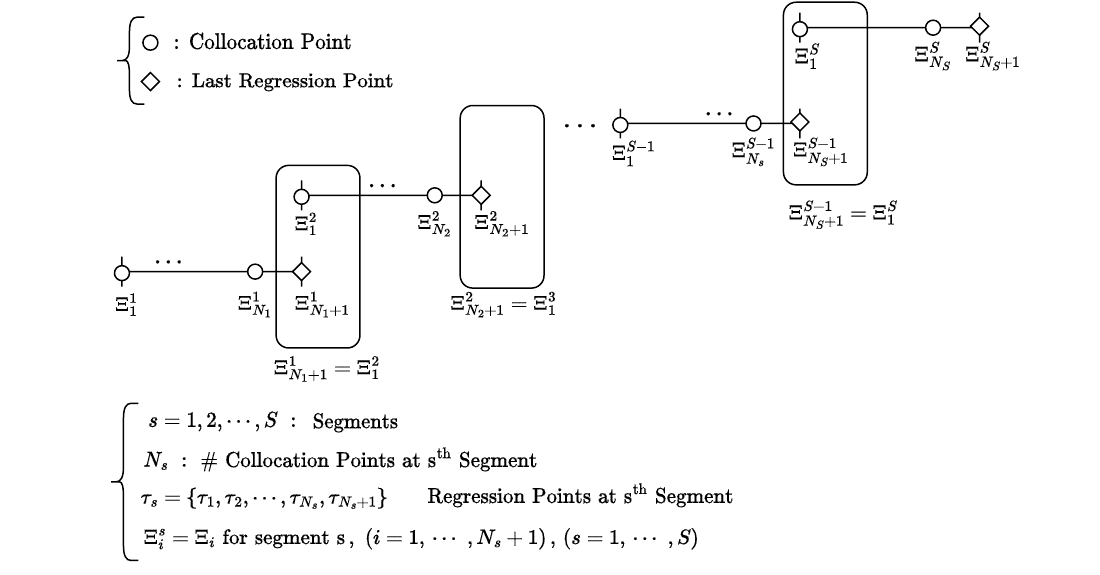}
    \caption{Applied constraint to satisfy continuity of states through segments}
    \label{fig:coll_points}
\end{figure*}

where $n_{\xi}$ shows number of states and .$n_{u}$ shows number of controls. As a result, we have:
\begin{align}
    \bm D_k \bm \Xi^{\mathrm{LGR}}=\frac{t_f-t_0}{2}\bm f\left( \bm \Xi_k, \bm U_k,\tau,t_o,t_f \right)\,,\, (k=1,\cdots,N)
\end{align}

Furthermore, the Lagrange part of the objective can be obtained through numerical quadrature. As a result, the equivalent NLP form of the OLC is:

\begin{equation}
\begin{aligned}
\min_{\bm \Xi^{\mathrm{LGR}},\,\bm U^{\mathrm{LGR}}}J &= \Phi(\bm \Xi(\tau_1),\tau_1,\bm \Xi(\tau_{N+1}),\tau_{N+1}) \\
& \quad + \frac{t_f-t_0}{2}\sum_{k=1}^{N} w_k L(\bm \Xi_k,\bm U_k, \tau,t_0,t_f)\\
    &\,\,\,\,\,\,\mathrm{Subject\,\,to}:\\
    \bm D_k \bm \Xi^{\mathrm{LGR}}-&\frac{t_f-t_0}{2}\bm f\left(\bm \Xi_k,\bm U_k,\tau,t_0,t_f \right)\,=\,0\\
    &\bm \phi \left( \bm \Xi(\tau_1),\tau_1,\bm \Xi(\tau_{N+1}),\tau_{N+1} \right)\,\,=\,\,0\\
    &\frac{t_f-t_0}{2}\bm{\mathcal{C}}\left( \bm \Xi_k,\bm U_k,\tau,t_0,t_f \right)\le \\
    &\mathrm{for\,\, } k=1,\cdots,N
\end{aligned}
\end{equation}

In cases where the time window $[t_0, t_f]$ is large, it can lead to a large number of collocation points and a correspondingly high-order polynomial, which can introduce difficulties. To address this, a strategy involves partitioning the time window into $S$ segments and employing an $N^{th}$ degree polynomial within each segment. Let there be $S$ segments, denoted by $s=1, \cdots, S$, with $N_s$ collocation points allocated to each segment. Additionally, a regression point (at $\tau^s_{N+1}=1$) is placed at the end of each segment. This segmentation introduces additional constraints: The terminal state value at the current segment must be equal to the initial state value of the succeeding segment. It's noteworthy that while the terminal value doesn't coincide with a collocation point, the initial value does, as depicted in Figure~\ref{fig:coll_points}.

\section{Implementation}
\label{sec:Imp}
\begin{figure}
\hspace*{-1.5cm}
    \centering
    \includegraphics[scale=0.7]{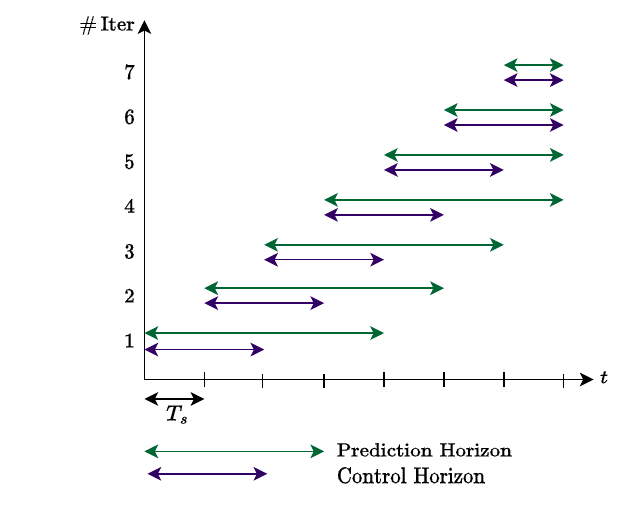}
    \caption{Prediction horizon and control horizon through mpc iterations}
    \label{fig:Horizon}
\end{figure}

One important difference between MPC and OLC is the time window. In OLC, the optimization problem is solved through the entire time window $[t_0,t_f]$. However, in MPC, the problem is solved many time. At each iteration, the time interval is $[t^{\mathrm{iter}}_0\,\,t^{\mathrm{iter}}_f]$ . Where, $t^{\mathrm{iter}}_0=(\mathrm{iter-1})T_s$, and $t^{\mathrm{iter}}_f=\min(t_f,t^{\mathrm{iter}}_0+pT_s)$. In addition, the time window of controller is $[t^{\mathrm{iter}}_{0c}\,\,t^{\mathrm{iter}}_{fc}]$ . Where, $t^{\mathrm{iter}}_{0c}=(\mathrm{iter-1})T_s$, and $t^{\mathrm{iter}}_{fc}=\min(t_f,t^{\mathrm{iter}}_{0c}+mT_s)$. Figure~\ref{fig:Horizon} shows prediction horizon and control horizon time window at each iteration. Here, it is assumed $m=2$, and $P=3$. As we see, the length of time window may decrease as it reaches the final time.

\begin{figure}
    \centering
    \includegraphics[width=0.99\linewidth]{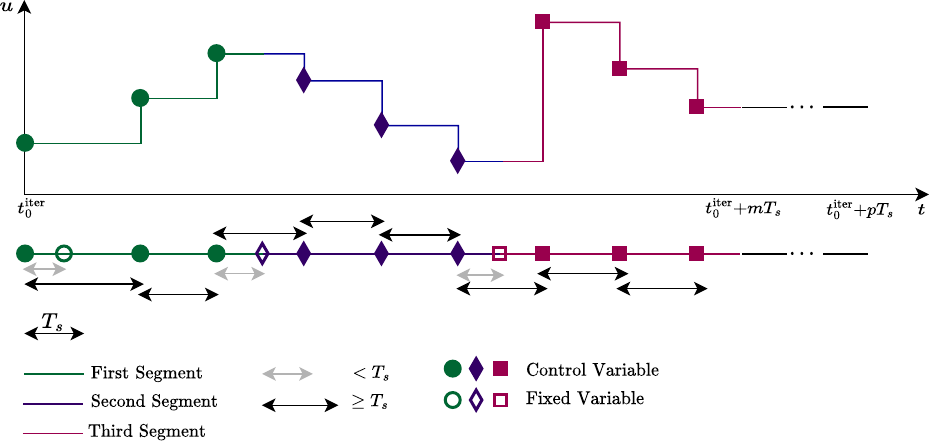}
    \caption{Control variables at each iteration that satisfy $T_s$ sampling rate}
    \label{fig:Ts}
\end{figure}

Another important difference between MPC and OLC is the sample rate, which is fixed between each sample point at $T_s$. As was mentioned before, the collocation points are obtained by LGR nodes, and the difference between two consecutive LGR nodes may be smaller than $T_S$. Therefore, the control at those points should not be considered as a design variable but should be fixed and equal to previos node that has satisfied these criteria. This concept is shown in Fig.~\ref{fig:Ts}. The final time of the control horizon is $(t^{\mathrm{iter}}_{0c}+mT_s)$, so as we see, the control at a time beyond this point is fixed and is equal to the control signal at this final point. In times smaller than $(t^{\mathrm{iter}}_{0c}+mT_s)$, this criteria is checked on collocation points. If the length of time window between two LGR points is less than $T_s$, it is not considered a control variable but is fixed. These points are shown by hollow shapes. Other points are shown by filled shapes. In addition, the control signal between these control points is constant and equal to the control variable at that point.

\begin{figure}
\hspace*{-1.5cm}
    \centering
    \includegraphics[width=1.0\linewidth]{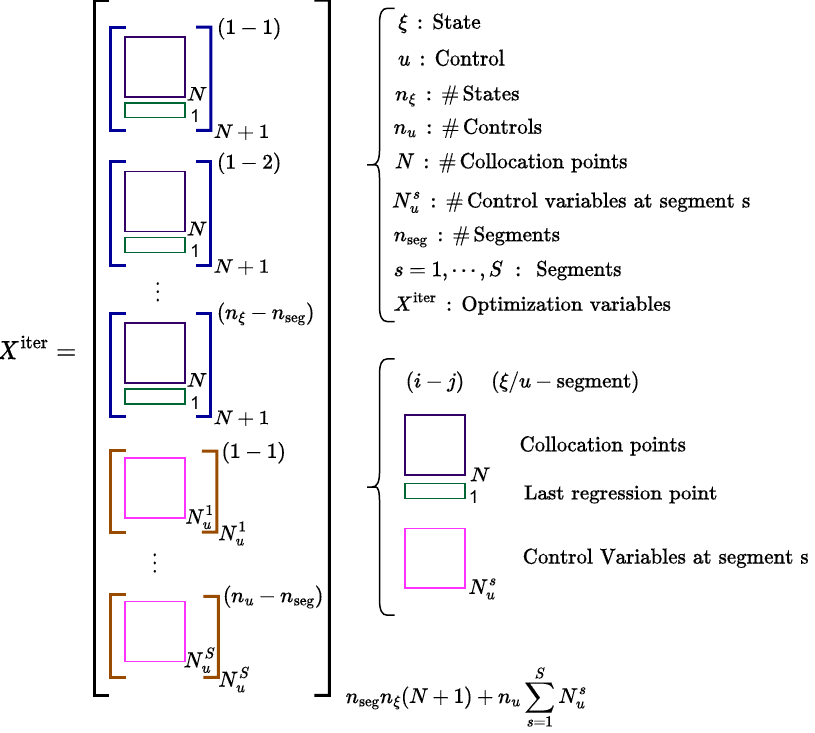}
    \caption{MPC variables at each iteration which includes both state and control values }
    \label{fig:X0_Iter}
\end{figure}

At each iteration, the optimization variables include states and controls. States are defined at LGR points and end points of each section, which are used for regression. However, for the control signal, those collocation points in the control time window that satisfy $T_s$ criteria are considered as control variables. Fig.~\ref{fig:X0_Iter} shows the optimization variable at each iteration. $\bm X^{\mathrm{iter}}$ consists of many blocks. Each block corresponds to state and control variables at each segment. 

\begin{figure}
\hspace*{-1.5cm}
    \centering
    \includegraphics[scale=0.7]{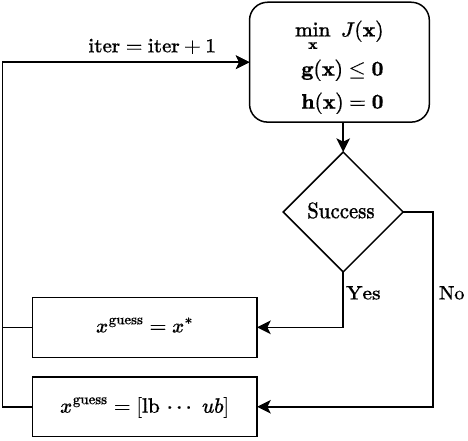}
    \caption{Method to provide initial guess}
    \label{fig:XIter_Flowchart}
\end{figure}

As a similar problem is solved at each iteration, the current result can be used as a guess for the next. This is shown in Fig.~\ref{fig:XIter_Flowchart}. If the success flag of the optimization algorithm is ``True'', then the result is considered as the initial guess. However, if it is not true, then a signal that varies linearly from lower bound to upper bound is considered for the initial guess. Additionally, the number of design variables may vary from iteration to iteration. The code also takes these into account, but they are not shown in this flowchart for simplicity. 

One thing that is important for optimization algorithms is scaling. Here, all state and control variables are scaled to $[-1,1]$. To do these, 4 matrices are considered in the code that should be defined by the user : $A_{\mathrm{scaled}},B_{\mathrm{scaled}},C_{\mathrm{scaled}},\,\mathrm{and}\,D_{\mathrm{scaled}}$. $A_{\mathrm{scaled}}\,\mathrm{and}\,B_{\mathrm{scaled}}$ are arrays with length $n_{\xi}$, and $C_{\mathrm{scaled}}\,\mathrm{and}\,D_{\mathrm{scaled}}$ are arrays with length $n_{u}$. $A_{\mathrm{scaled}}(i)=\frac{\mathrm{ub}^i+\mathrm{lb}^i}{2}$, $B_{\mathrm{scaled}}(i)=\frac{\mathrm{ub}^i-\mathrm{lb}^i}{2}$, $C_{\mathrm{scaled}}(j)=\frac{\mathrm{ub}^j+\mathrm{lb}^j}{2}$, $D_{\mathrm{scaled}}(j)=\frac{\mathrm{ub}^j-\mathrm{lb}^j}{2}$, for $i=1,\cdots\,n_\xi$, and $j=1,\cdots,n_u$, where ``lb'' is the lower bound and ``ub'' is upper bound.

One more thing that is used in the code is ODE simulation. When pseudo-spectral methods are used for optimal control, feasibility is one of the biggest concerns. I.e., the initial mesh may not be adequate to satisfy dynamic equations and regridding is needed. Here, after each iteration, the dynamic is simulated in that time interval using obtained control. As a result, the result is always feasible.

\section{Software Structure}
\label{sec:Sof-Struct}

In this section, the different structures that need to be defined by the user are defined, and different parts of the code are shown in Fig.~\ref{fig:Code_struct}. Here we go through them one by one:

\begin{figure}[t]
    \centering
    \subcaptionbox{}{\includegraphics[scale=0.6]{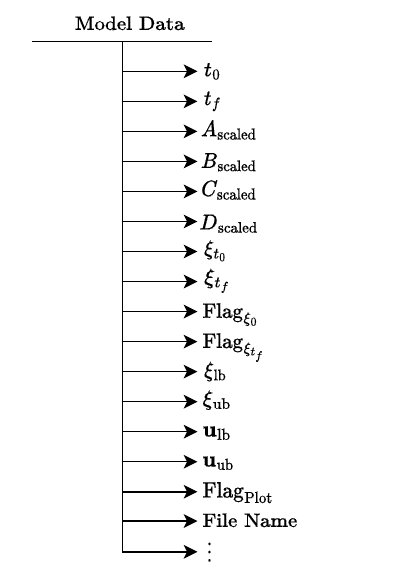}}
    \subcaptionbox{}{\includegraphics[scale=0.6]{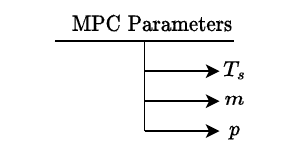}}
    \subcaptionbox{}{\includegraphics[scale=0.6]{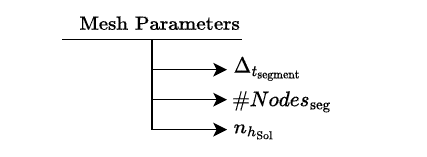}}
    \subcaptionbox{}{\includegraphics[scale=0.6]{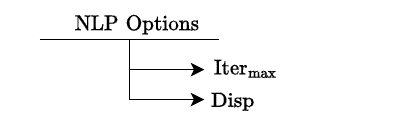}}
    \subcaptionbox{}{\includegraphics[scale=0.6]{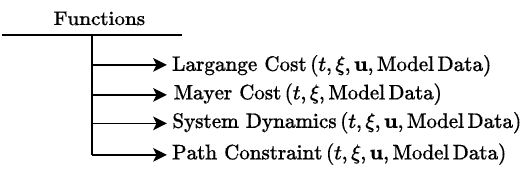}}
    \caption{Different structures used in MPC software}
    \label{fig:Code_struct}
\end{figure}

\begin{itemize}
    \item Model Data:\\
    $t_0$ is initial time and $t_f$ is final time. Scaling matrices are A, B, C, and D. $\bm \xi_{t_0}$ and $\bm \xi_{t_f}$ are states at the initial time and final time and are used as boundary constraints. $\mathrm{Flag}_{\bm \xi_0}$ and $\mathrm{Flag}_{\bm \xi_f}$ are flags, which shows whether boundary constraints should be used at the initial and final point or not. $\bm {\xi}_{\mathrm{lb}}$ and $\bm {\xi}_{\mathrm{ub}}$ are lower bound and upper bound of states, and $\bm {u}_{\mathrm{lb}}$ and $\bm {u}_{\mathrm{ub}}$ are lower bound and upper bound of controls. $\mathrm{Flag}_{\mathrm{Plot}}$ is a flag that shows the plot at each iteration if the flag is ``True''. However, this increases the computation time. $\mathrm{File\,\,Name}$ is an address to save the result. All other data that are needed in system dynamics, constraints, or objectives should also be defined in the Model Data Structure. The Model Data is input to dynamics, constraints, and objective functions so that the user can get access to these data through the Model Data structure.
     \item MPC Parameters:\\
     $T_s$ is control sampling time, $m$ is control horizon, and $p$ is prediction horizon.
     \item Mesh Parameters:\\
     As was mentioned before, at each iteration, MPC is divided into several sections and each section has $N$ nodes. Here $\Delta_{t_{\mathrm{segment}}}$ shows time interval for each section. And  $\# \mathrm{Nodes}_{\mathrm{seg}}$ shows number of nodes at each section. As a result, at each iteration the time interval is divided into some sections base on the division of that time window to $\Delta_{t_{\mathrm{segment}}}$ and $\# \mathrm{Nodes}_{\mathrm{seg}}$ shows number of nodes at each segment. $n_{h_{\mathrm{sol}}}$ is an integer value and time step that is used to store ODE result is $\frac{T_s}{n_{h_{\mathrm{sol}}}}$.
     \item NLP Options:\\
     $\mathrm{Iter}_{\mathrm{max}}$ is maximum number of iterations. $\mathrm{Disp}$ is a flag and if it is ``True'', optimization algorithm displays optimization messages. Here Scipy is used; therefore, Scipy messages will be displaced.
     \item Functions:\\ This structure includes 4 functions: Lagrange cost, Mayer cost, system dynamics, and path constraints. The input of all these functions are time, state, control, and Model Data. The output of Lagrange cost is $\mathcal{L}$. the output of Mayer cost is $\mathcal{M}$, the output of system dynamics is $\bm f$ and the output of path constraint is $\mathcal{C(\cdot)}>0$ which is a vector and if all elements are equal or grater than zero then the path constraint is satisfied.
\end{itemize}

\begin{figure}
    \centering
    \includegraphics[width=1.0\linewidth]{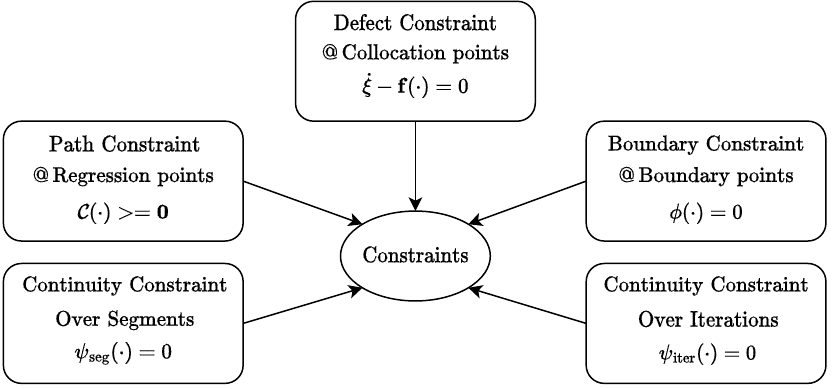}
    \caption{Constraint Flowchart}
    \label{fig:Const_Flowchart}
\end{figure}

Different constraints used in this code is shown in Fig.~\ref{fig:Const_Flowchart}. These are defined as:
\begin{itemize}
    \item Defect Constraint:\\
    This satisfies the dynamic equation at collocation points. The user needs to define dynamics, and this code itself used that function to generate defect constraints at all collocation points.
    \item Path Constraint:\\ This uses path constraint function to provide path constraint.
    \item Boundary Constraint:\\ This uses $\bm \xi_{t_0}$, $\bm \xi_{t_f}$, $\mathrm{Flag}_{\xi_{t_0}}$ and $\mathrm{Flag}_{\xi_{t_f}}$ to generate boundary constraints.
    \item $\Psi_{\mathrm{iter}}$ provide continuity equations over iterations. In other words, the starting value of current iteration is equal to end value of last iteration.
    \item $\Psi_{\mathrm{seg}}$ provide continuity equations over segments. In other words, the regression point of current segment is equal to first collocation point of next segment.
\end{itemize}

\begin{figure}
    \centering
    \includegraphics[width=1.0\linewidth]{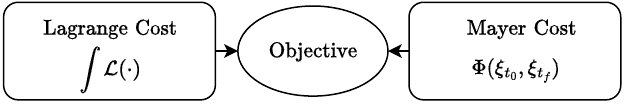}
    \caption{Objective Flowchart}
    \label{fig:Objective_Flowchart}
\end{figure}

Different objective functions used in this code are shown in Fig.~\ref{fig:Objective_Flowchart}. These are defined as:

\begin{itemize}
    \item Lagrange cost: This function computes Lagrange cost based on user Lagrange cost defined in Functions.
    \item Mayer cost: This function uses user Mayer cost defined in Functions to compute Mayer cost.
\end{itemize}

The whole algorithm is shown in the Algorithm.~\ref{fig: QL-Alg}. First, it reads user-defined data and functions. Then it computes all LGR and regression points, goes through each iteration, and solves the MPC problem. Then it uses the ODE solver and stores the data, and based on that, it provides an initial point for states for the next iteration. This is posed as a constraint. In the end, all data are stored in a pickle file whose name is provided by the user, and the results are plotted.

\begin{algorithm}
\caption{Software Procedure}\label{fig: QL-Alg}
\textbf{Train}\;
$\mathrm{Define\,Model\,\,Data}$\;
$\mathrm{Define\,MPC\,\,Parameters}$\;
$\mathrm{Define\,Mesh\,\,Parameters}$\;
$\mathrm{Define\,dynamics,\,constraint,\,and \,obj \,funcs}$\;
$\mathrm{Define\,Scipy\,Options}$\;
 \For{$\mathrm{iter}$ $\mathrm{in}$ range($n_{\mathrm{iter}}$)}{
 $t_0^{\mathrm{iter}}=(\mathrm{iter}-1)T_s$\;
 $t_f^{\mathrm{iter}}=\min(t_f,t_0^{\mathrm{iter}}+pT_s)$\;
 $t_{f_c}^{\mathrm{iter}}=\min(t_f,t_0^{\mathrm{iter}}+mT_s)$\;
  \If{$\mathrm{Flag}_{\mathrm{Plot}}==\mathrm{True}$}{$\mathrm{Plot\,\,Results \,\,in \,\,[t_0^{\mathrm{iter}},\,t_f^{\mathrm{iter}}]}$}
 $\xi^{\ast}, u^{\ast} \gets \mathrm{Solve\,\,MPC\,\,in\,\,interval\,[t_0^{\mathrm{iter}},\,t_f^{\mathrm{iter}}]}$\;
 $\mathrm{Store\,\,Data} \gets \mathrm{Run \,ODE\,\,Dynamics\,\,in\,\,[t_0^{\mathrm{iter}},\,t_{f_c}^{\mathrm{iter}}]}$\;
$\mathrm{Update\,\xi_0\,for\,the\,next\,iterartion\,uinsg \,ODE}$
 }
 $\mathrm{load\,Data\,\,in\,\,pickle\,\,file}$\;
 $\mathrm{Plot\,Result\,\,in\,\,[t_0,\,t_f]}$\;
\end{algorithm} 

\section{Example}
\label{sec:Examples}

In this section, an example with an analytical solution will be explored, and a comparison will be drawn between the analytical solution and the one derived through MPC. More exmaples are studied in .\ref{sec:Examples-appendix}. In these examples the effect of control sampling time, control horizon, and prediction horizon are investigated. Also the code in the \href{https://github.com/saeidb71/A-User-Friendly-Software-Based-on-Legendre-Gauss-Radau-Pseudo-Spectral-Method-in-Python-to-Solve-MPC.git}{\underline{GitHub repository}} provides a good examples how the user should define problems.

The problem studied here is obtained from \citet{bryson2018applied} pp.166-167:

\begin{align}
    &\min_{\bm u(t)}  \frac{1}{2}\int^{t_f}_{t_0} u^2 dt\\
    &\mathrm{where:}\\
    &\dot{\bm \xi}=\begin{bmatrix} 
                    0 & 1\\
                    -1 & 0
                    \end{bmatrix}
                    \bm \xi +
                    \begin{bmatrix} 
                    0\\
                    1
                    \end{bmatrix} u\\
    & \xi_1(0)=x_0,\,\, \xi_2(0)=v_0,\,\, \xi_1(t_f)=0,\,\, \xi_2(t_f)=0              
\end{align}

The exact open-loop optimal control when $x_0=-1/2$ and $v_0=1$ is:
\begin{align}
\scriptsize u^{\ast}(t) = -\frac{2}{t_f^2-\sin^2(t_f)}
\scriptsize
\begin{bmatrix}
x_0\\
v_0
\end{bmatrix}^T
\begin{bmatrix}
\sin(t_f-t)\sin(t_f)-t_f\sin(t)\\
-\cos(t_f-t)\sin(t_f)+t_f\cos(t)
\end{bmatrix}
\end{align}

\begin{figure}[ht!]
    \centering
    \subcaptionbox{}{\includegraphics[scale=0.5]{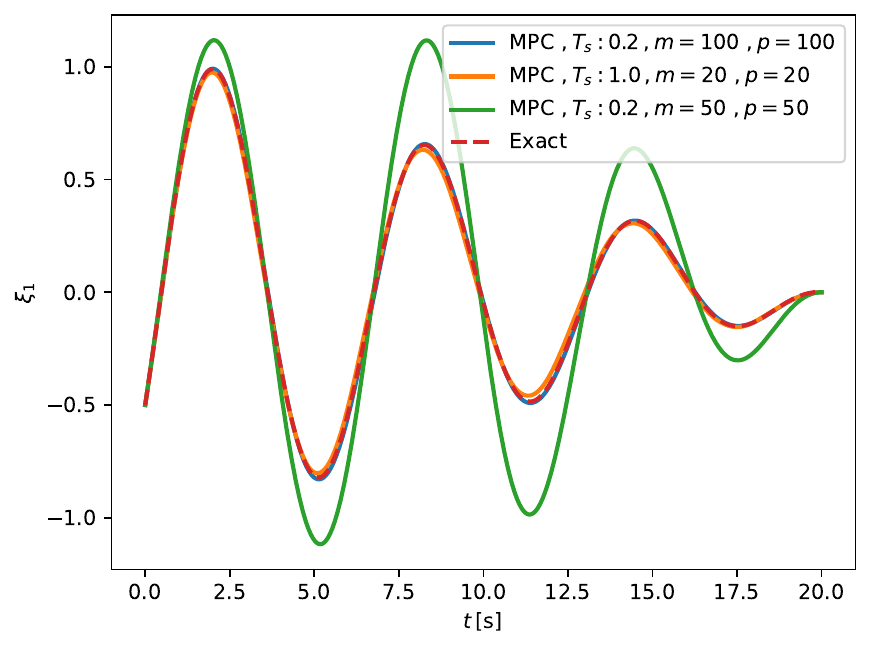}}
    \subcaptionbox{}{\includegraphics[scale=0.5]{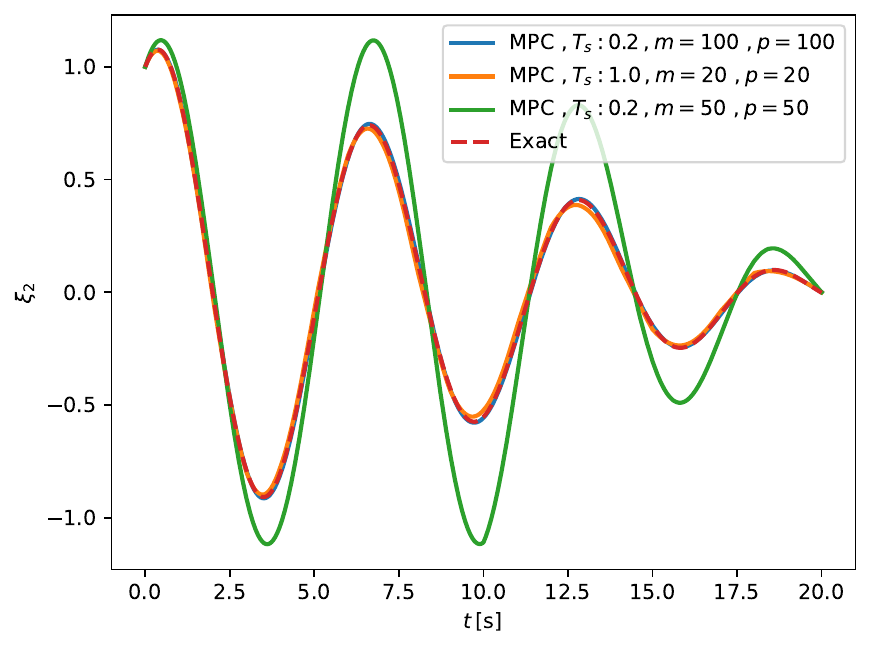}}
    \subcaptionbox{}{\includegraphics[scale=0.5]{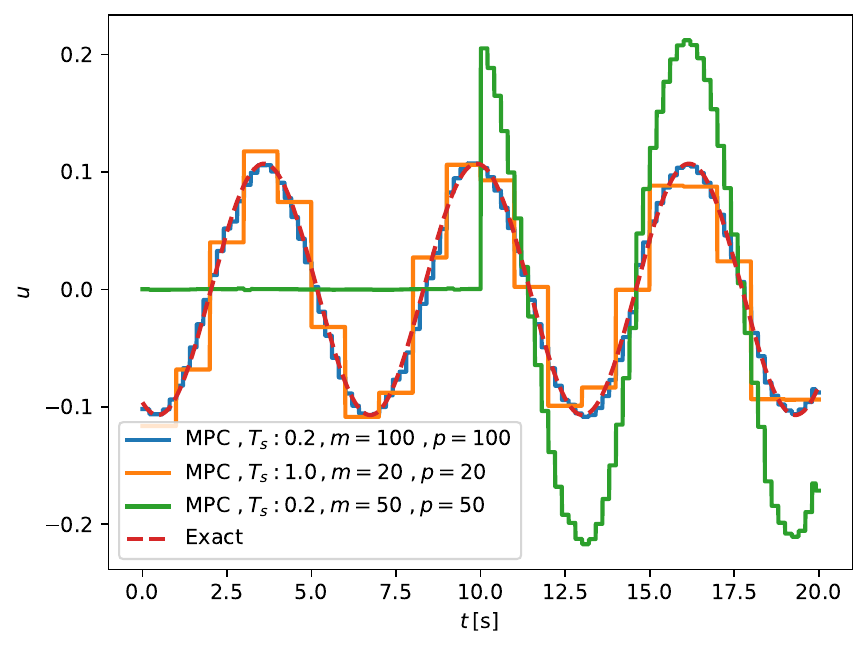}}
    \caption{Examples 1 response through different scenarios. Here ``exact'' is obtained by analytical solution, and other legends show the corresponding MPC parameters in each scenario.}
    \label{fig:Ex1}
\end{figure}

The analytical solution and MPC results are shown in Fig.~\ref{fig:Ex1}. In the first two MPC cases, the prediction horizon is the same as the final time ($t_f$), so it has access to the full time-window, and the only difference between these two cases is in their sampling time:  In the first one, $T_s=0.2$, and in the second one $T_s=1.0$. In the third case, the time window is $1$ s which is smaller than the final time ($2$ s), so the MPC has no information about boundary constraints at the final point until the start time of the MPC is 1 s. As it is shown in Fig.~\ref{fig:Ex1}(c), the control signal is 0 from start to $1$ s, and after that, by having access to the boundary constraint, the control changes to satisfy that constraint. 

\begin{figure}[ht!]
    \centering
    \subcaptionbox{}{\includegraphics[scale=0.5]{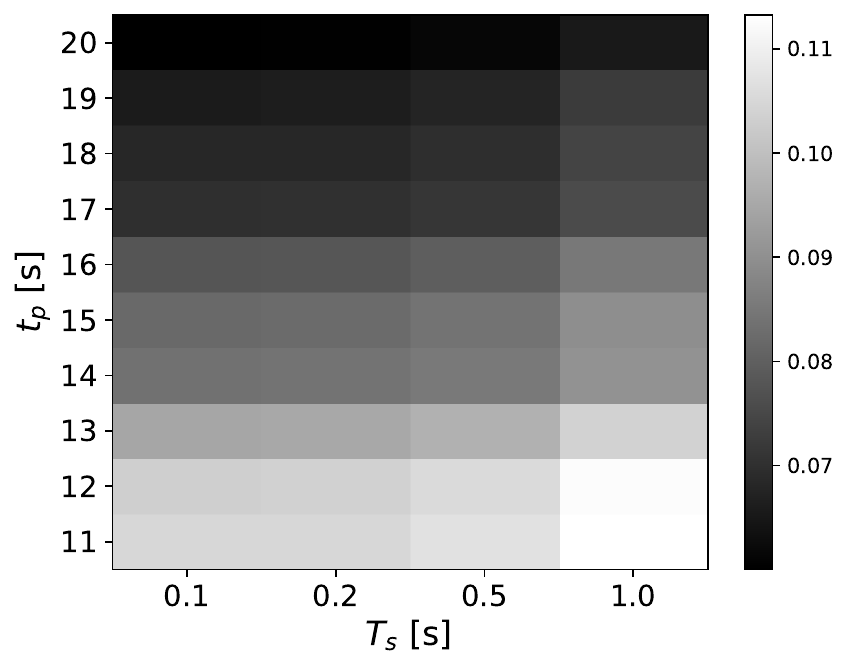}}
    \subcaptionbox{}{\includegraphics[scale=0.5]{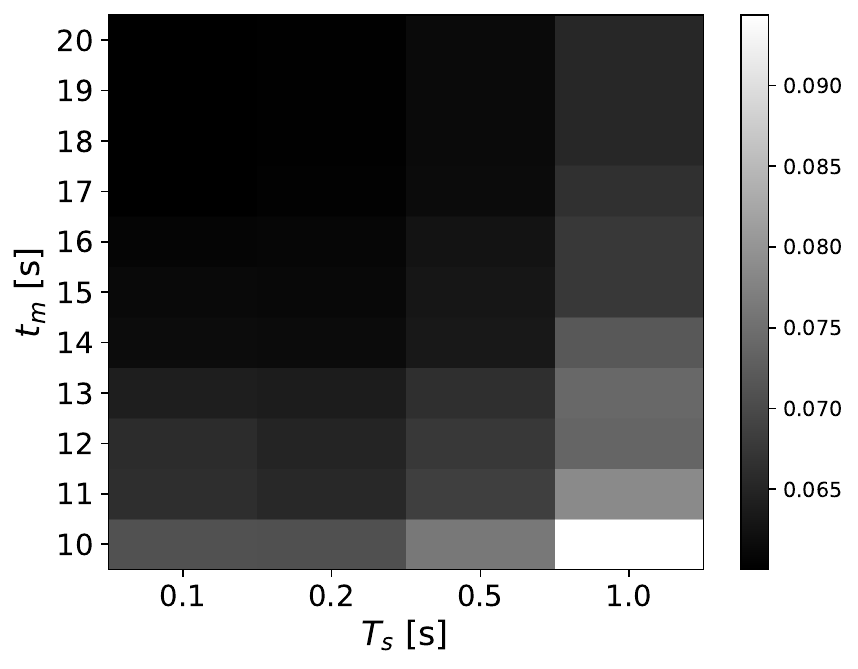}}
    \caption{(a) Exploring objective values across a 2D design space of $T_s$ and $t_p$, assuming $t_m=t_p$. (b) Exploring objective values across a 2D design space of $T_s$ and $t_m$, assuming $t_p=20.0 \mathrm{s}$}
    \label{fig:Ex1_sweep}
\end{figure}

Using this software, it is possible to modify MPC parameters such as $T_s$, $t_p$, and $t_m$, enabling an exploration of their impact on the outcomes. In Fig.~\ref{fig:Ex1_sweep}(a), the variation of the objective value across the 2D space defined by $T_s$ and $t_p$ is presented. In this case, the assumption is that $t_m$ is equal to $t_p$. Evidently, shifting towards the upper left corner of the plot leads to diminished objective values, signifying an improved outcome. This outcome arises because a greater authority is vested in the MPC controller in this region. Additionally, Fig.~\ref{fig:Ex1_sweep}(b) displays the objective value across the 2D domain of $T_s$ and $t_m$, assuming that $t_p=20$. Once again, moving towards the upper left corner results in enhanced performance. This enhancement is due to the reduction in $T_s$ and the increase in $t_m$ along this trajectory. These analyses play a crucial role in guiding decisions related to actuator, hardware, and sensor selection. For instance, if there's no need for excessively small $T_s$, more cost-effective actuators and hardware components can be adopted. Alternatively, if variations in $t_m$ have minimal influence on the outcomes, economical sensors can be preferred.

\section{Conclusion}
\label{sec:Concl}

In this article, a user-friendly software is developed to solve MPC problems. This problem is converted to a nonlinear program using the pseudo-spectral method, and the optimization problem is solved using Scipy. This code is written in python, and the corresponding \href{https://github.com/saeidb71/A-User-Friendly-Software-Based-on-Legendre-Gauss-Radau-Pseudo-Spectral-Method-in-Python-to-Solve-MPC.git}{\underline{Github address}} is provided to get access to the source code and examples. This code was tested under different examples with exact solutions, and the MPC results were compared with the exact one.  This paper demonstrated how this code works and how different parts should be defined. Due to its minimal requirement for in-depth MPC understanding, engineers from diverse disciplines can employ it effectively. 



\section*{Declaration of Competing Interest}

The authors declare that they have no competing interest.

\section*{Data Availability}

All data required to replicate the results can be generated by the Python optimization code. The Python optimization codes for all the problems demonstrated in the manuscript are available online using the mentioned GitHub repository.

\appendix

\section{Methods to solvey MPC}
\label{sec:Methods}

\begin{enumerate}
    \item Indirect Method: This method is also called ``optimize then discretize'' where Pontryagin's Maximum Principle (PMP) or Calculus of Variation (CV) are used to derive optimality equations \citep{rao2009survey} , and then numerical methods are used to discretize the problem and solve it.  The Strength and shortcomings of this method are:
    \begin{enumerate}
        \item [$+$]: Derived optimality equations gives us insight into the structure of the solution, ex: Linear Quadratic Regulator (LQR) 
         \item [$-$]: Obtained boundary value problem is difficult to solve
         \item [$-$]: Path constraints are difficult to handle
         \item [$-$]: Initial guess for co-state is needed, and this is not straightforward because co-states do not represent physical states
         \item [$-$]: Problem is ill-conditioned and is not robust
         \item [$-$]: Cannot be used for black-box functions
    \end{enumerate}
    \item Direct Methods: These methods are also called ``discretize then optimize'' because the problem is first discretized and then solved using an NLP solver \citep{fahroo2008advances}. The Strength and shortcomings of this method are:
    \begin{enumerate}
        \item  [$+$]: Much more effective than direct methods
        \item  [$+$]: More effective in handling constraints
        \item  [$+$]: Powerful NLP solver can be used
        \item  [$+$]: There exist well-established methods
        \item  [$-$]: Mathematics is not as elegant as indirect methods
        \item  [$-$]: It gives us no insight into the structure of the problem
    \end{enumerate}
    Direct methods includes 2 sub-methods: ``Sequanrtial'' and ``Simultaneous'':
    \begin{enumerate}
        \item Sequential: In this method, just the control signal is discretized and dynamic is satisfied through simulation, and the result is always feasible \citep{herber2014dynamic}. As a result, the dynamic equation in the optimal control problem is not under the ``Subject to'' section but will be in the ``where'' section because it is no longer satisfied by the optimization algorithm but is obtained through simulation. The strength and shortcomings of this method are:
        \begin{enumerate}
            \item  [$+$]: Smaller number of defect constraints than simultaneous because states are not discretized
            \item  [$+$]: Feasible solution can be easily find
            \item  [$+$]: Powerful differential algebraic solvers can be used 
            \item  [$-$]: Need to perform full simulation for each parameter perturbation
            \item  [$-$]: Is not effective for highly nonlinear problems or unstable problems
            \item  [$-$]: It is computationally inefficient
        \end{enumerate}
        The sequential method itself includes two methods: Single Shooting and Multiple Shooting \citep{herber2017advances}:
        \begin{enumerate}
            \item Single Shooting: In this method, the whole prediction horizon time window is simulated in one section. The strength and shortcomings of this method are:
            \begin{enumerate}
                \item  [$+$]: Simple
                \item  [$+$]: Dynamics are always feasible
                \item  [$+$]: No defect constraint is needed
                \item  [$-$]: NLP inherits the ill-conditioning of the problem
            \end{enumerate}
            \item Multiple Shooting: In this method, the time window is divided into several sections, and each section is optimized independently. The states at these section points must be identical and satisfied through defect constraints. I.e., the state and final time of the previous section will be the same as the state at the initial time of the current section. The strength and shortcomings of this method are:
            \begin{enumerate}
                \item [$+$]: Better convergence than single shooting
                \item [$+$]: problem is well-conditioned
                \item [$+$]: it has a good structure for parallel computation
                \item [$-$]: More complex than single shooting
                \item [$-$]: Defect constraint is needed so make the NLP more complex
            \end{enumerate}
        \end{enumerate}
        
         \item Simultaneous: In this method, both control and states are discretized. In this approach, the dynamic equation is satisfied through defect constraints. The strength and shortcomings of this method are:
         \begin{enumerate}
             \item  [$+$]: The resulting NLP is sparse, so powerful sparse NLP solvers like IPOPR and SNOPT can be used
             \item  [$+$]: Knowledge of the state trajectory can be used as a guess in the initialization
             \item  [$+$]: It has fast local convergence
             \item  [$+$]: Effective for unstable systems
             \item  [$+$]: Effective for path and boundary constraints
             \item  [$+$]: Work well even for ill-conditioned problems because the NLP does not inherit the ill-conditioning of the problem
             \item  [$-$]: a large number of defect constraints
             \item  [$-$]: Mesh regrinding is needed and changes the NLP dimension
             \item  [$-$]: More challenging to find a feasible solution than Sequential metho
             \item  [$-$]: State derivative function ($\bm f$) is needed, so it does not work with data-driven models
         \end{enumerate}
         The simultaneous method itself consists of two methods: Single Step and Pseudo Spectral method:
         \begin{enumerate}
             \item Single Step: In this method, the location of discretized points is selected evenly, and low-order polynomials are used in each section. The strength and shortcomings of this method are:
             \begin{enumerate}
                 \item [$+$] Less difficult to implement than the Pseudospectral method
                 \item [$-$] Finer mesh than pseudospectral needs to be sued
                 \item [$-$] more computation cost than pseudospectral
             \end{enumerate}
             \item Pseudospectral: In this method, the location of discretized points are not placed evenly but are obtained through roots of some polynomials like Legendre polynomials. Furthermore, high-order polynomials like Lagrange polynomials are used to interpolate states. The strength and shortcomings of this method are:
             \begin{enumerate}
                 \item [$+$] Coarser mesh than single step can be used
                 \item [$+$] Lead to an easier problem for NLP than single step
             \end{enumerate}
         \end{enumerate}
    \end{enumerate}
\end{enumerate}

\section{Examples}
\label{sec:Examples-appendix}

\subsection{Example 2}
\label{sec:sub-Example2}
The second problem is obtained from \citet{bryson2018applied} pp.120-122. 

\begin{align}
    &\min_{\bm u(t)} \int^{t_f}_{t_0} \frac{1}{2}u^2 dt \label{eq: ex2}\\
    &\mathrm{where:}\nonumber\\
    &\dot{\bm \xi}=\begin{bmatrix} 
                    0 & 1\\
                    0 & 0
                    \end{bmatrix}
                    \bm \xi +
                    \begin{bmatrix} 
                    0\nonumber\\
                    1
                    \end{bmatrix} u\nonumber\\
    & \xi_1(0)=0,\,\, \xi_2(0)=v_0,\,\, \xi_1(t_f)=1,\,\, \xi_2(t_f)=-1\nonumber\\
    &\xi_1(t)\le l\nonumber
\end{align}

In this example, there are two states and one control. Based on the bounds and constraints shown in Eq.~\ref{eq: ex2}, the scaled matrices can be defined as:

\begin{align}
    A_{\mathrm{scaled}}&=\begin{bmatrix}
    (l-l)/2\\
    (2-2)/2
    \end{bmatrix} \\
    B_{\mathrm{scaled}}&=\begin{bmatrix}
    (l+l)/2\\
    (2+2)/2
    \end{bmatrix} \\
    C_{\mathrm{scaled}}&=\begin{bmatrix}
    (20-20)/2
    \end{bmatrix} \\
    D_{\mathrm{scaled}}&=\begin{bmatrix}
    (20+20)/2
    \end{bmatrix} 
\end{align}

The exact open-loop optimal control when $l=1/9$ is:
\begin{equation}
    u^{\ast} = \left\{ \begin{array}{ll}
         -\frac{2}{3l}\left( 1-\frac{t}{3l} \right) & \mbox{if $0 \leq t \le 3l$};\\
         0 & \mbox{if $3l \leq t \le 1-3l$};\\
        -\frac{2}{3l}\left( 1-\frac{1-t}{3l} \right) & \mbox{if $1-3l \leq t$}.\end{array} \right.
\end{equation}

\begin{figure}[ht!]
    \centering
    \subcaptionbox{}{\includegraphics[scale=0.5]{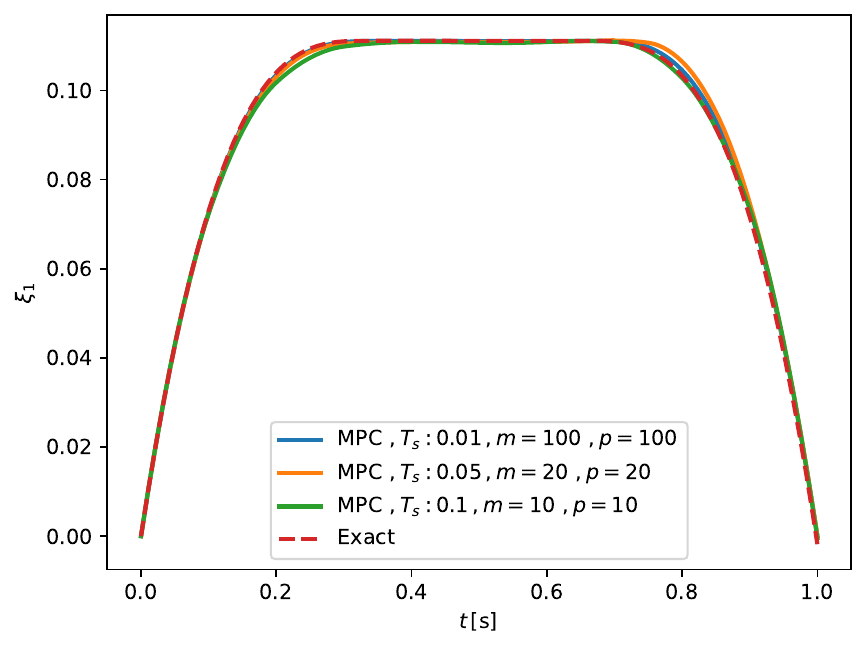}}
    \subcaptionbox{}{\includegraphics[scale=0.5]{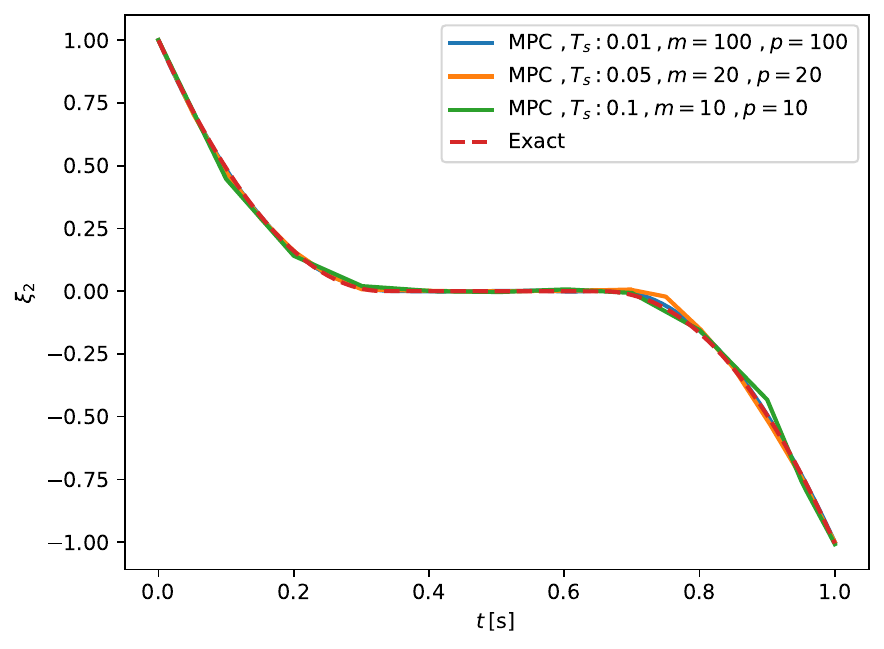}}
    \subcaptionbox{}{\includegraphics[scale=0.5]{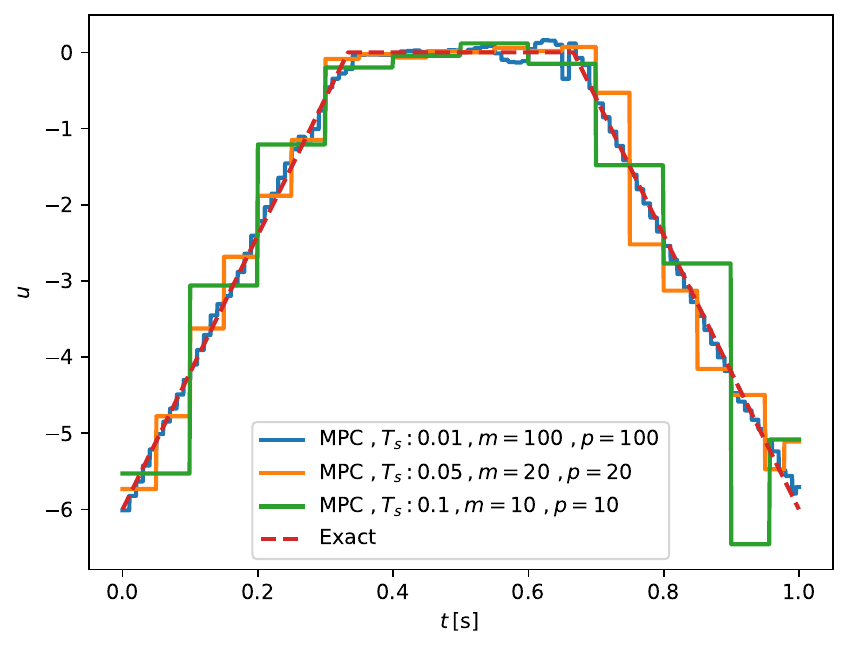}}
    \caption{Examples 2 response through different scenarios. Here ``exact'' is obtained by analytical solution, and other legends show the corresponding MPC parameters in each scenario.}
    \label{fig:Ex2}
\end{figure}

The analytical solution and MPC results are shown in Fig.~\ref{fig:Ex2}. In all 3 MPC cases, the prediction horizon is the same as the final time, and only the control sampling time is different. The result is closer to the analytical solution when the control sampling time is shorter. When it is bigger, the result will deviate from the exact solution obtained by open-loop optimal control.

\subsection{Example 3}
\label{sec:sub-Example3}
The third problem is obtained from \citet{bryson2018applied} pp.109-110. 

\begin{align}
    \min_{\bm u(t)} \frac{a^2}{2}\xi^2_{t_f}+& \int^{t_f}_{t_0} \frac{1}{2}u^2 dt\\
    &\mathrm{where:}\\
    &\dot{\bm \xi}=b(t)u\\
    & \xi(0)=\xi_0\\
    &|u(t)|\le 1
\end{align}

The exact open-loop optimal control when is:
\begin{equation}
    u^{\ast} = -\mathrm{sat}\left[ a^2b(t)\xi(t_f) \right]
\end{equation}

\begin{figure}[ht!]
    \centering
    \subcaptionbox{}{\includegraphics[scale=0.5]{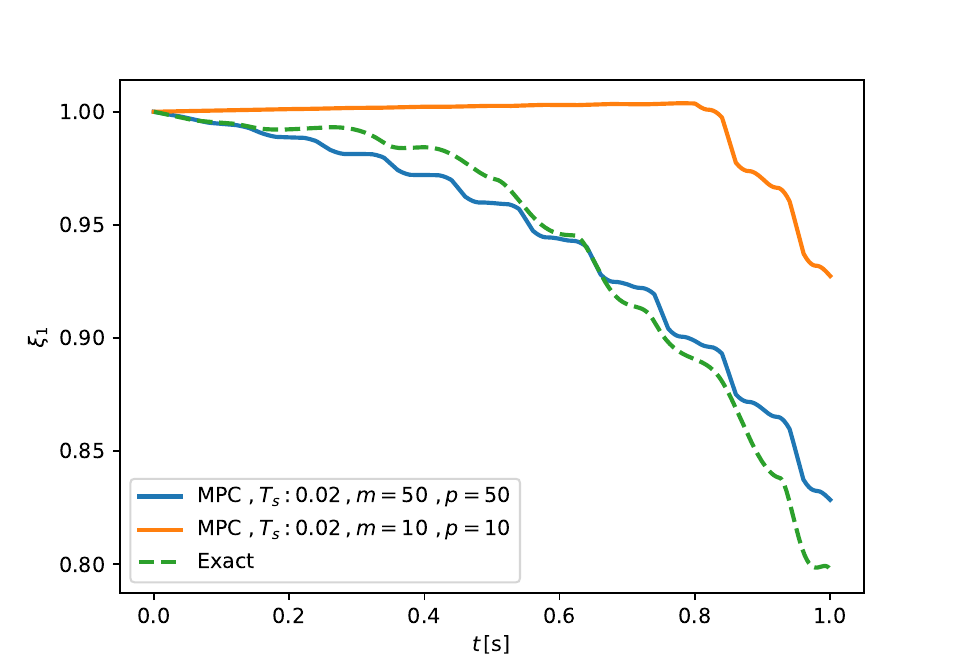}}
    \subcaptionbox{}{\includegraphics[scale=0.5]{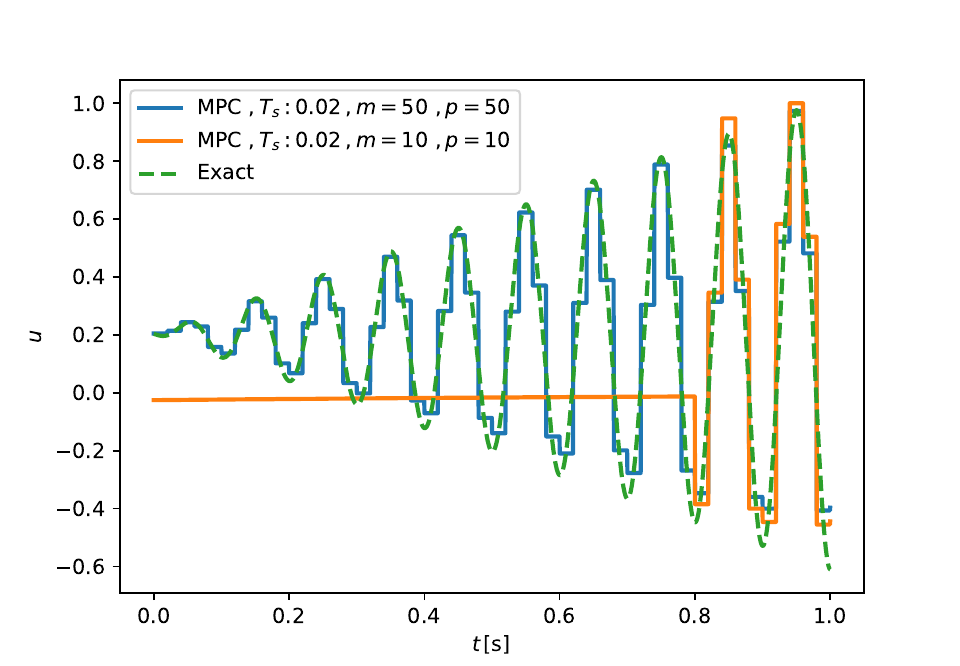}}
    \caption{Examples 3 response through different scenarios. Here ``exact'' is obtained by analytical solution, and other legends show the corresponding MPC parameters in each scenario.}
    \label{fig:Ex3}
\end{figure}

The analytical solution and MPC results are shown in Fig.~\ref{fig:Ex3}  for the case where $t_f=1$, $\xi_0=1$, $a=1$ and $b(t)=t\cos(20 \pi t)-1/4$. Here control sampling time is the same for both MPC cases, and the only difference is in their prediction horizon. In the first MPC case, the prediction horizon covers the full time-window, but in the second case, it only covers $0.2$ s., which is smaller than the final time ($1$ s); as a result, in this case, MPC has no access to the Mayer cost, until it reaches $0.8$ s. As we see, the control is zero for this case from $t=0$ to $t=0.8$, while for the first case, the result is close to the exact solution.

\subsection{Example 4}
\label{sec:sub-Example4}

\begin{figure}[ht!]
\hspace*{-1.5cm}
    \centering
    \includegraphics[scale=0.4]{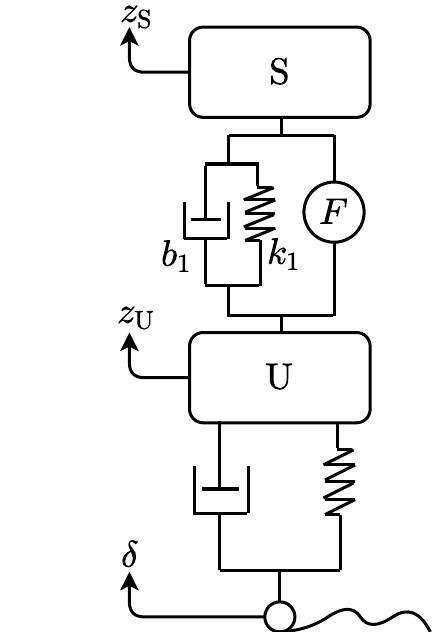}
    \caption{Vehicle suspension system}
    \label{fig:VHC}
\end{figure}

Fig.~\ref{fig:VHC} shows a simple model of vehicle suspension system defined in  \citet{herber2019problem, SaeidBayat-Vehicle}. Here $\delta$ is road profile, $U$ is unsprung mass, $S$ is sprung mass, $z_\mathrm{U}$ is the displacement of unsprung mass, and $z_\mathrm{S}$ is the displacement of the sprung mass. Furthermore, $F$  shows control force, $m$ shows mass, $k$ shows spring, and $b$ shows damper. The goal is to provide a smooth ride for passengers by transferring forces through different components like spring, damper, mass, and actuator. To reach this goal, the objectives is defined in Eq.~\ref{eq: obj} \citep{herber2019problem}. Here $w_1(z_U-\delta)^2$ represents  handling performance, $w_2\Ddot{z}_S^2$ shows passenger comfort, and $w_3F^2$ is a penalty function for control effort. In this example, the initial value of states is set to zero, and the displacement between sprung and unsprung mass is bound by $r_\mathrm{max}$. The parameters used in this example are whin in Table.~\ref{table: Vh_params}. For the MPC control sampling time, prediction horizon and control horizon are:  $T_s=0.01$, $p=50$, and $m=50$, so the length of time window interval is $0.5$ s.

\begin{figure*}[ht!]
    \centering
    \subcaptionbox{}{\includegraphics[scale=0.46]{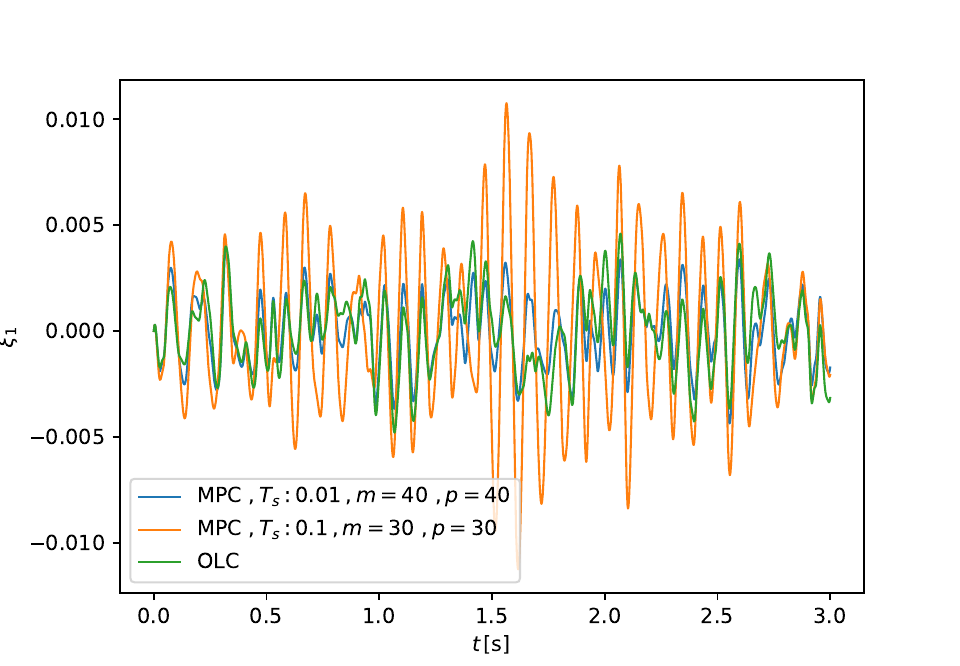}}
    \subcaptionbox{}{\includegraphics[scale=0.46]{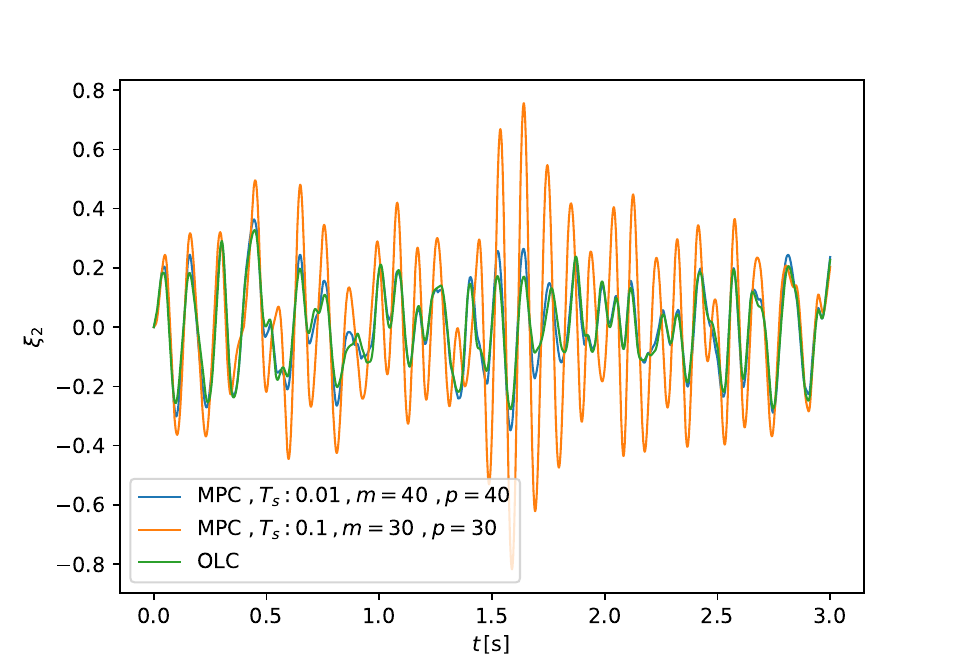}}
    \subcaptionbox{}{\includegraphics[scale=0.46]{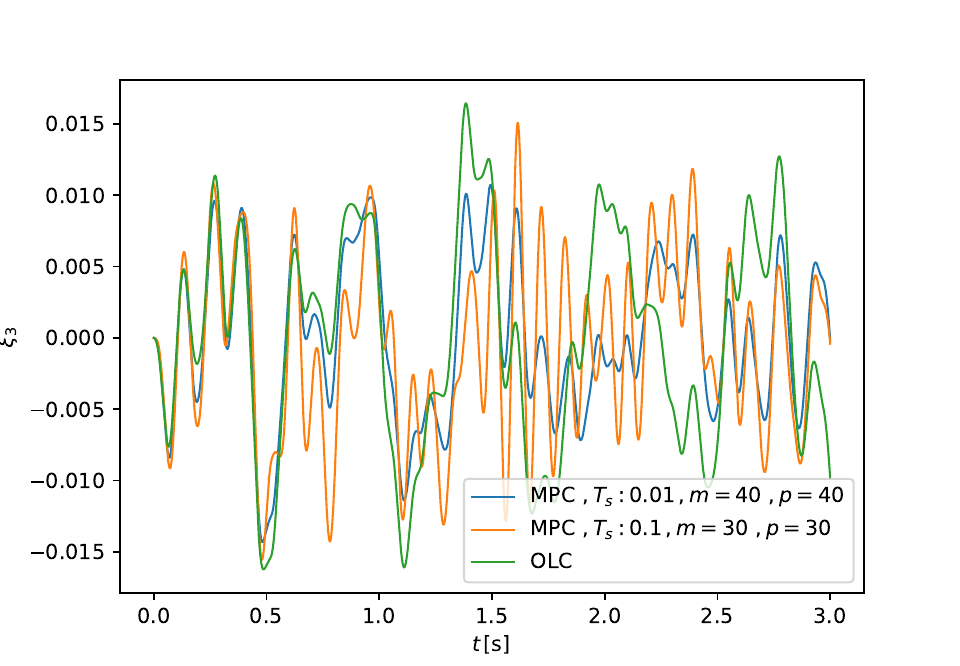}}
    \subcaptionbox{}{\includegraphics[scale=0.46]{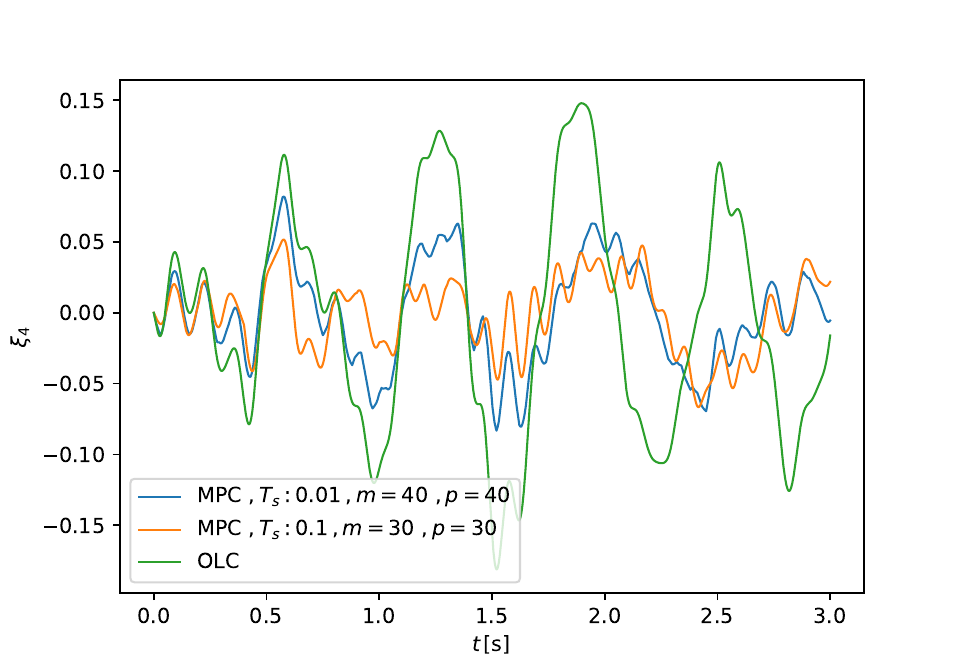}}
    \subcaptionbox{}{\includegraphics[scale=0.46]{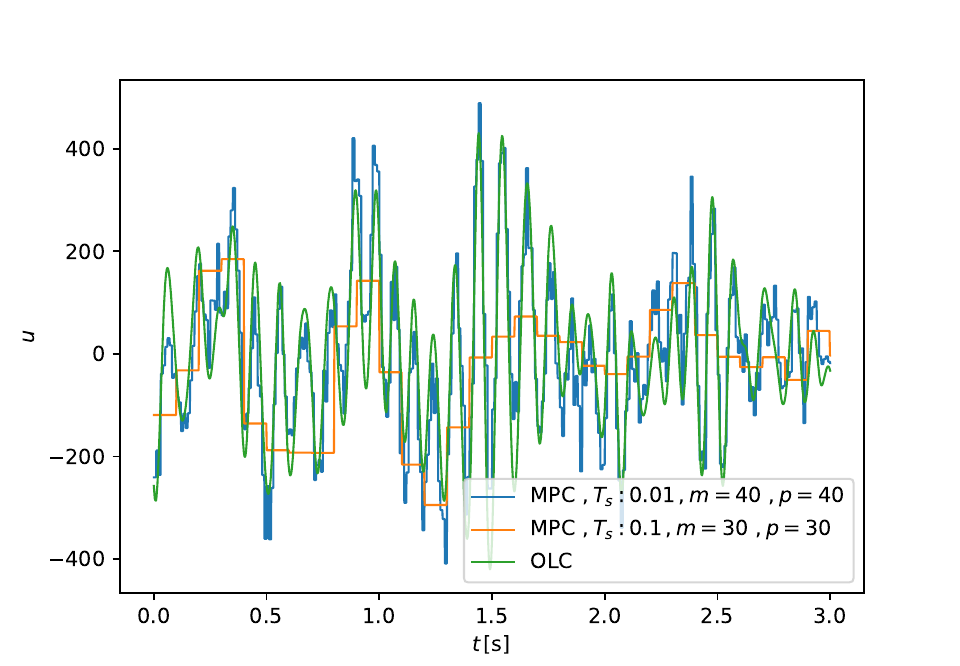}}
    \caption{Vehicle suspension results}
    \label{fig:Ex4}
\end{figure*}

Figure.~\ref{fig:Ex4} shows the results obtained by MPC and OLC. Here, the entire time horizon is from $0$ to $3$ seconds. Here $\xi_1=z_U-\delta$,\,$\xi_2=\dot{z}_U$,\,$\xi_3=z_S-z_U$,\, $\xi_4=\dot{z}_S$. In the first MPC case the prediction horizon is $0.4\, \mathrm{[s]}$ and control sampling time is $0.01\, \mathrm{[s]}$, and in the second MPC case, the prediction horizon is $3\, \mathrm{[s]}$ and control sampling time is $0.1\, \mathrm{[s]}$. As we see, the result of the first case is much closer to the OLC because of having a shorter sampling time. The OLC response has a high frequency, so here having a shorter sampling time is more important than having a bigger prediction horizon.

\begin{align}
\label{eq: obj}
    \Uppsi_d=\int_{t_0}^{t_F} \left( w_1(z_U-\delta)^2+w_2\Ddot{z}_S^2+w_3F^2\right)\mathrm{d}t
\end{align}

\begin{table}[ht!]
\centering
\caption{Vehicle suepnsion parameters}
\label{table: Vh_params}
\resizebox{\linewidth}{!}{ \begin{tabular}{cc|cc}
\toprule
Parameter & Value & Parameter & Value \\ \hline
    $t_0$      &    0 s   &       $t_f$    &     3 s   \\
    $b$      &   $100.8$ Ns/m    &        $k$   &   $2.15*10^4$ N/m    \\
   $r_{\mathrm{max}}$       &   $0.04$ m    &        $s_{\mathrm{max}}$    &     $0.04$ m   \\
   $m_{\mathrm{U}}$       &   $65$ kg    &          $m_{\mathrm{S}}$ &    $325$ kg   \\
   $w_1$       &   $10^5\, s^{-1}m^{-2}$    &       $k_t$    &  $232.5\times10^3$ N/m     \\
   $w_2$          &  $0.5\, s^{3}m^{-2}$     &      $b_t$     &    $0$ Ns/m   \\
   $w_3$       &  $10^{-5}\, s^{-1}N^{-2}$     &           &       \\ \bottomrule
\end{tabular}}
\end{table}

\bibliographystyle{elsarticle-num-names} 
\bibliography{Main}
\end{document}